# Fermion number conservation isn't fermion conservation[*]

Wolfgang Bock[a], James E. Hetrick[b] and Jan Smit[b]

[a] University of California, San Diego, Department of Physics,
9500 Gilman Drive 0319, La Jolla, CA 92093-0319, USA

[b] Institute of Theoretical Physics, University of Amsterdam,
Valckenierstraat 65, 1018 XE Amsterdam, The Netherlands

A nonperturbative regularization of the Standard Model may have a superficially undesirable exact global U(1) symmetry corresponding to exact fermion number conservation. We argue that such a formulation can still have the desired physics of fermion nonconservation, i.e. fermion particle creation and annihilation by sphaleron transitions. We illustrate our reasoning in massless axial QED in 1+1 dimensions.

The Standard Model has fermion number violation, through sphaleron transitions in the anomaly in the $B + L$ current. A typical form of a lattice fermion action, $S_F = -\overline{\psi} M(A) \psi$, with $M(A)$ the fermion matrix, has an exact global U(1) invariance $\psi \to \exp(i\omega)\psi$, $\overline{\psi} \to \exp(-i\omega)\overline{\psi}$. Assuming that the lattice fermion measure in the path integral is also U(1) invariant, one expects this invariance to be accompanied by exact fermion number conservation. This suggests that such a lattice model cannot possibly represent the Standard Model.

In a customary way of thinking about this, the lattice model should violate explicitly the global U(1) invariance, and this violation should turn into the anomaly in the scaling region [1].

We shall present here a different possibility: in a theory with exact fermion number conservation we can still have fermion particle creation and annihilation by sphaleron transitions. The crucial observation is that fermions are excitations relative to the surface of the Dirac sea, and although a state cannot change its fermion number, the Dirac sea, being the ground state with lowest energy, can do so.

A simple model for discussing these issues is massless axial QED in 1+1 dimensions, with

$$S_F = -\int dt\, dx\, \overline{\psi}\gamma^\mu(\partial_\mu + iA_\mu \gamma_5)\psi. \qquad (1)$$

[*]Presented by J. Smit

By a charge conjugation on e.g. the right handed component of the fermion fields this theory is equivalent to massless (vector) $QED_2$. In the continuum lore, the axial model has an anomaly in the divergence equation of its vector current, while the vector model has an anomaly in the divergence equation of its axial current.

We can learn about fermion creation by studying spectral flow in an external gauge field that changes slowly with time. The idea is that we consider time slices of instanton-like configurations, starting with a vacuum gauge field at early times, going through a sphaleron at time zero, and ending up at later times with a different vacuum gauge field. We can then determine if an initial fermion vacuum state has evolved into a state with particles, produced by the gauge field in accordance with the anomaly. If the change with time is very slow, we can deduce the evolution of the states by continuity, by following the eigenstates of the time dependent hamiltonian (adiabatic theorem).

We take space to be a circle with circumference $L$ and choose the gauge specified by $A_t(x, t) = 0$ and $A_x(x, t) = A(t)$. Then the time dependence is expressed by $A$-dependence. In this gauge, the Chern-Simons number $C = AL/2\pi$. Going through an instanton-like configuration means that $C$ changes by one unit. Values $A = 0 \pmod{2\pi/L}$ are 'pure gauge', since for these values we can write $A = \Omega i \partial_x \Omega^*$, with $\Omega = \exp(i2\pi kx/L)$ a periodic gauge transformation with winding number $k$.

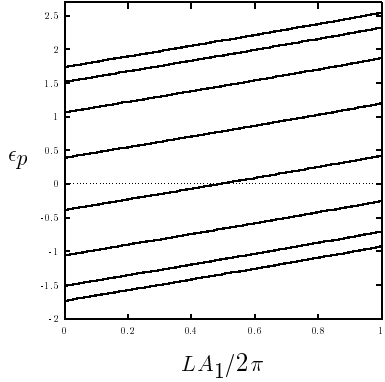

Figure 1. Spectral flow in the staggered fermion axial QED$_2$ model.

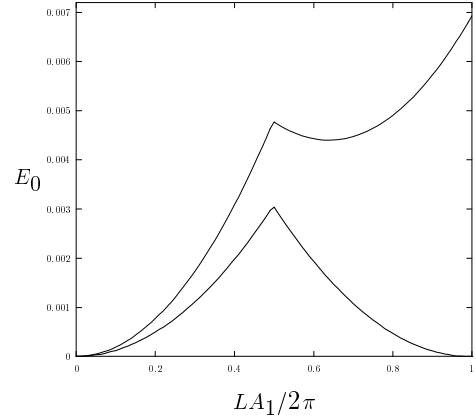

Figure 2. Ground state energy in the staggered axial QED$_2$ model, before (upper curve) and after (lower curve) addition of the $A^2$ counterterm.

The ground state $|0, A\rangle$ of the hamiltonian $H(A)$ is by definition the state with lowest energy. When $A$ is pure gauge, these ground states are to be identified with the vacuum. In Hilbert space, gauge transformations on $A$ are represented by unitary transformations, and we have to identify all states related by gauge transformations, hence also all vacua with differing integer Chern-Simons numbers.

We study the spectral flow from $A = 0$ to $A = 2\pi/L$, for which $C$ changes from 0 to 1, and are interested in the state $|\Psi, A\rangle$, which starts out as the vacuum $|0, 0\rangle$ (i.e. $|\Psi, 0\rangle = |0, 0\rangle$) and 'flows' into the final state $|\Psi, 2\pi/L\rangle$. We can then compare $|\Psi, 2\pi/L\rangle$ with the final vacuum $|0, 2\pi/L\rangle$.

It is instructive to look first at the continuum theory with a sharp momentum cutoff $\Lambda$. Then the axial QED hamiltonian $H$ and the equivalent vector QED hamiltonian $H'$ are given by

$$H = \sum_p [\psi_R^\dagger \psi_R(p+A) + \psi_L^\dagger \psi_L(-p+A)]$$
$$\quad - AN/2,$$
$$H' = \sum_p [\psi_R'^\dagger \psi_R'(p-A) + \psi_L^\dagger \psi_L(-p+A)],$$

where the summation is over momenta $p = (2n+1)\pi/L \in [-\Lambda, \Lambda]$, $\sum_p \equiv (1/L)\sum_n$, and $N$ is the total number of modes. $L$ and $R$ denote the left and right handed projections, and we suppressed the momentum label on $\psi(p) = \int dx \exp(-ipx)\psi(x)$. The charge conjugation relating the axial and vector model is given by $\psi_R' = \psi_R^\dagger$, $\psi_R'^\dagger = \psi_R$; $H'$ has the same spectrum as $H$. There are in these models with sharp momentum cutoff regularization two conserved charges: both $F = \sum_p [\psi_R^\dagger(p)\psi_R(p) + \psi_L^\dagger(p)\psi_L(p)] - N/2$, and $F_5 = \sum_p [\psi_R^\dagger(p)\psi_R(p) - \psi_L^\dagger(p)\psi_L(p)]$ commute with $H$. In the vector version the two charges are given by $F' = -F_5$, $F_5' = -F$.

The spectral flow of the eigenmode energies $\epsilon$ in the axial model is very simple, $\epsilon_p = p + A$, for $R$ and $\epsilon_p = -p + A$ for $L$, so each mode energy increases linearly with $A$. In the vector version $\epsilon_p' = \pm(p - A)$, and the $R$ ($L$) modes decrease (increase) linearly with $A$.

Consider now the flow of the state $|\Psi, A\rangle$, which starts out as the vacuum state at $A = 0$, hence with all negative energy modes occupied. Each mode is doubly degenerate ($L$ and $R$). In the axial model the $\epsilon_p$ flow upwards with $A$ and at $A = 2\pi/L$ they have taken the place of their predecessor, except for the two modes starting out as $\epsilon = -\pi/L$ and ending up at $\epsilon = +\pi/L$. The quantum numbers $F$ and $F_5$ are conserved, both are zero for all states $|\Psi, A\rangle$, as for the initial vacuum state $|0, 0\rangle$. The energy of the state $|\Psi, A\rangle$ also happens to be constant in with this regularization. However, the ground state $|0, A\rangle$ looses two occupied modes half way at $A = \pi/L$, when the $\epsilon = -\pi/L + A$ mode gets above zero. So the final ground state energy $E_0$ differs from the inital one. This is possible because of the gauge noninvariance of the sharp momentum cutoff. The important point is now that the final ground state has two occupied states ($L$ and $R$) less than the initial ground state, so the final vacuum quantum numbers are $F = -2$ and $F_5 = 0$. Hence, $\Delta F \equiv (F_\Psi - F_{vac})_{\text{final}} - (F_\Psi - F_{vac})_{\text{initial}} = +2$, just as expected from the anomaly in the vector current, $\Delta F = 2\Delta C$. In the vector model the $L$ modes move upwards and the $R$ modes move downwards. Midway, the ground state looses an $L$ mode and gains an $R$ mode, such that $\Delta F' = 0$, $\Delta F'_5 = -2$.

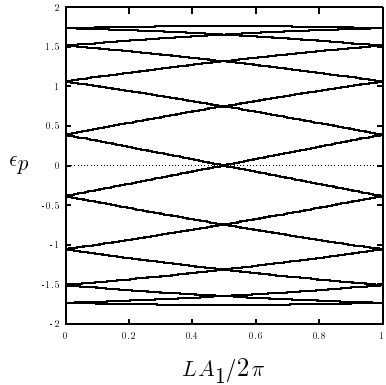

Figure 3. Spectral flow in the staggered fermion vector $QED_2$ model.

In the axial model we have the creation of two particles ($L$ and $R$), whereas in the vector model we have the creation of a particle ($L$) and an antiparticle (also $L$, an $R$-hole).

The continuum model with a sharp momentum cutoff is non-local and therefore unsatisfactory. However staggered fermion models on the lattice are local, with otherwise similar properties. Our lattice formulation of axial QED will be the 'canonical' staggered fermion model described in [2]. This model violates axial gauge invariance, but a simple mass counterterm for the gauge field restores invariance in the scaling region, for smooth external gauge fields. The model has indeed an exact global U(1) symmetry corresponding to fermion number conservation. (It has also an even number of 'flavors', but this is not important for our discussion.) We have calculated the flow of the energy eigenvalue spectrum as defined by the transfer operator.

Fig 1 shows spectral flow of the axial QED model on a lattice with 16 sites in space. We see some lattice artefacts in the level spacing near the cutoff. The ground state energy is shown in fig. 2 (32 sites). It is very different from the continuum model, but also here the lack of gauge invariance is clear. However, the addition of a local counterterm in the action $\propto \sum_{x,t} A_\mu A_\mu$ restores gauge invariance, as seen in fig. 2.

In the vector theory we can of course have exact gauge invariance on the lattice. Fig. 3 shows the spectral flow in the usual staggered fermion $QED_2$ for a lattice with 16 sites in space. The initial and final spectra coincide, in accordance with gauge invariance. It appears that the modes at the cutoff switch $L \leftrightarrow R$. This is possible because the 'staggered $\gamma_5$' behaves only like $\gamma_5$ for momenta in the scaling region. Near the cutoff the labels '$L$' and '$R$' lose their usual meaning. The ground state energy in this gauge invariant vector model is indistinguishable from the lower curve in fig. 2. The curvature $E_0 \approx LA^2/\pi$ in the minima of $E_0$ determines the mass of the boson in the Schwinger model.

We expect a similar mechanism to work also in realistic four dimensional models, including Yukawa couplings to Higgs fields.

Acknowledgements. We would like to thank J. Vink for useful conversations. This work is financially supported by the Stichting voor Fundamenteel Onderzoek (FOM) and by the DOE under contract DE-FG03-90ER40546.